\documentclass{article}
\usepackage{spconf,amsmath,graphicx}
\usepackage{subcaption}
\usepackage{tabularx}
\usepackage{comment}

\title{Using uncertainty estimation to reduce false positives in liver lesion detection}
 \name{Ishaan Bhat$^{\dagger}$ \qquad Hugo J. Kuijf$^{\dagger}$  \qquad Veronika Cheplygina$^{\star}$ \qquad Josien P.W. Pluim$^{\dagger \star}$}
 \address{$^{\dagger}$ Image Sciences Institute, University Medical Center Utrecht, The Netherlands\\
     $^{\star}$Department of Biomedical Engineering, Eindhoven University of Technology, The Netherlands}
\begin{document}
%
\maketitle
\begin{abstract}

Despite the successes of deep learning techniques at detecting objects in medical images, false positive detections occur which may hinder an accurate diagnosis. We propose a technique to reduce false positive detections made by a neural network using a SVM classifier trained with features derived from the uncertainty map of the neural network prediction. We demonstrate the effectiveness of this method for the detection of liver lesions on a dataset of abdominal MR images. We find that the use of a dropout rate of 0.5 produces the least number of false positives in the neural network predictions and the trained classifier filters out approximately $90\%$ of these false positives detections in the test-set.

\end{abstract}
\section{Introduction}
\label{sec:intro}

Primary tumors such as neuroendocrine and colorectal tumors have a high likelihood of developing metastases in the liver. Early detection of (new) liver metastases is crucial since it may prolong patient life~\cite{robinson_early_2002}. Automatic detection of these metastases is a challenging task and deep learning based systems are increasingly being used to address the challenge.

However, deep learning systems may make erroneous predictions. These arise due to a variety of reasons, for example, the model overfitting to the training data, presence of noise/artefacts in the image etc. Presence of false positives in the prediction is one such type of error and may hinder accurate patient diagnosis. 

Efficient and scalable uncertainty estimation techniques for deep learning-based systems such has MC-Dropout~\cite{gal_dropout_2015} and model ensembles~\cite{lakshminarayanan_simple_2017} have been widely adopted by the medical imaging research community to estimate uncertainty at tasks such as classification and segmentation. False positive detections tend to have a higher estimated uncertainty, thus uncertainty quantification can be used to filter such detections~\cite{nair_exploring_2018, mehrtash_confidence_2019, roy_inherent_2018}

There has been work to show that modern neural networks exhibit poor calibration~\cite{guo_calibration_2017} which may degrade the quality of uncertainty estimates~\cite{ovadia_can_2019}, thereby making any conclusion drawn on the basis of solely the uncertainty estimate, unreliable. In this paper, we propose an approach based on leveraging features based on shape and other attributes in addition to the uncertainty estimate to detect false positive predictions made by deep learning systems. 

\section{Related Work}
\label{sec:rel_work}
There has been active research to address challenges in developing interpretable uncertainty metrics to detect segmentation failures and aid clinicians in their decision making~\cite{roy_inherent_2018, jungo_uncertainty-driven_2018, nair_exploring_2018, sander_towards_2019}. In the context of image segmentation, computing such a metric on an entire object, rather than on a per-voxel basis may aid interpretability.The direct use of voxel-wise uncertainty estimates to detect failures has shown limited success~\cite{jungo_analyzing_2020}. In ~\cite{jungo_analyzing_2020}, it is also shown that aggregating voxel-wise uncertainties spatially can aid in detecting failed segmentations. 

In \cite{nair_exploring_2018} lesion-level uncertainties are computed by taking a log-sum of voxel-wise uncertainties over the lesion prediction by assuming that per-voxel uncertainty estimates within a single lesion volume are independent. It is shown that using lesion-level uncertainties to filter predicted lesions reduces the number of false positives and false negatives. In ~\cite{roy_inherent_2018, mehrtash_confidence_2019} a negative correlation between mean uncertainty over structure and the Dice score is shown, leading to the conclusion that the mean entropy over the structure can be used to filter wrong predictions. Similarly, in \cite{jungo_uncertainty-driven_2018} a \emph{doubt} score is computed by summing up voxel-wise uncertainties in predicted foreground regions.

An alternate approach to explicit aggregation of voxel-wise estimates has been to use a second neural network that uses the per-voxel uncertainty map and the network prediction to estimate the segmentation quality or refine detection~\cite{ozdemir_propagating_2017, devries_leveraging_2018}. In \cite{devries_leveraging_2018} a second neural network supplied with the prediction and the spatial uncertainty map learns to predict the Dice score. In \cite{ozdemir_propagating_2017} the second neural network uses a 3-channel input of the original image patch, prediction and uncertainty estimate to predict if the detection of the nodule by the first stage was correct.

In this paper, we propose a two-stage process to detect false positive predictions. Instead of training a second neural network, we train a SVM classifier to predict whether a lesion detected by the segmentation network is a false positive. This classifier is trained by computing a low-dimensional feature vector for each lesion, comprised of the aggregated uncertainty and shape-based attributes. Our approach requires less data to train the second stage (compared to the use of a neural network) and we demonstrate the effectiveness of this approach in Section \ref{sec:fpc}. 

\section{Methodology}
\subsection{Data}
In this paper we included abdominal DCE and DWI MRI of 72 patients with liver metastases from the University Medical Center Utrecht, the Netherlands.  

The DCE-MR series was acquired in six breath holds resulting in a total of 16 3-D images per patient. These images are used as a multi-channel input for the neural network. Voxel size for these images is $1.543$  x $1.543$  x $2$ mm . The liver and the metastases within the liver were manually segmented on the DCE-MRI by a radiologist in training and verified by a radiologist with more than 10 years of experience. The dataset mainly included colorectal metastases, neuroendocrine metastases and some other metastases types. The DCE-MR images were motion corrected using techniques presented in \cite{jansen_evaluation_2017}.

The DWI-MR images were acquired with three b-values: $10$, $150$, and $1000$ s/mm2, using a protocol with the following parameters: TE: $70$ ms; TR: $1.660$ ms; flip angle: $90$ degrees. For each patient, the DWI MR image was nonlinearly registered to the DCE MR image using the elastix\footnote{https://elastix.lumc.nl/} toolbox.

We apply the manually created liver masks to the abdominal DCE and DWI MR images and pre-process them using z-score normalization of the intensities. 

The data was split into 50 training patients, 5 validation patients and 17 test patients. 

\subsection{Neural Network Architecture and Training}
\begin{figure*}[htb]
  \centering 
  \includegraphics[width=15.0cm]{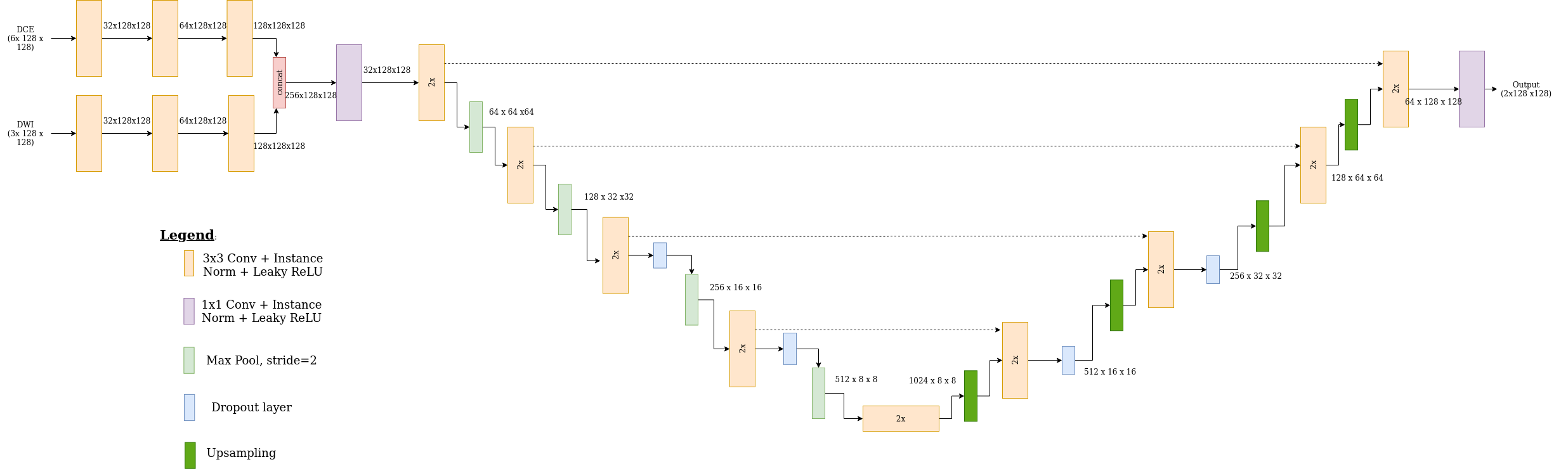}
  \caption{Neural network architecture used to perform lesion detection and uncertainty estimation}
  \label{fig:nn_arch}
\end{figure*}

Our choice of neural network architecture (Figure \ref{fig:nn_arch}) is inspired by the U-Net~\cite{ronneberger_u-net_2015} and Bayesian SegNet~\cite{kendall_bayesian_2016}. We use the standard encoder-decoder with skip connections like the U-Net and add dropout to the bottom-most encoder and decoder blocks since these posititions are shown to be most effective~\cite{kendall_bayesian_2016}. Preceding the encoder-decoder structures, we use convolutions to process and fuse the multi-channel DCE and DWI images.   

\begin{figure*}[htb]
\centering
\begin{subfigure}[t]{0.3\linewidth}
\includegraphics[height=4.0cm]{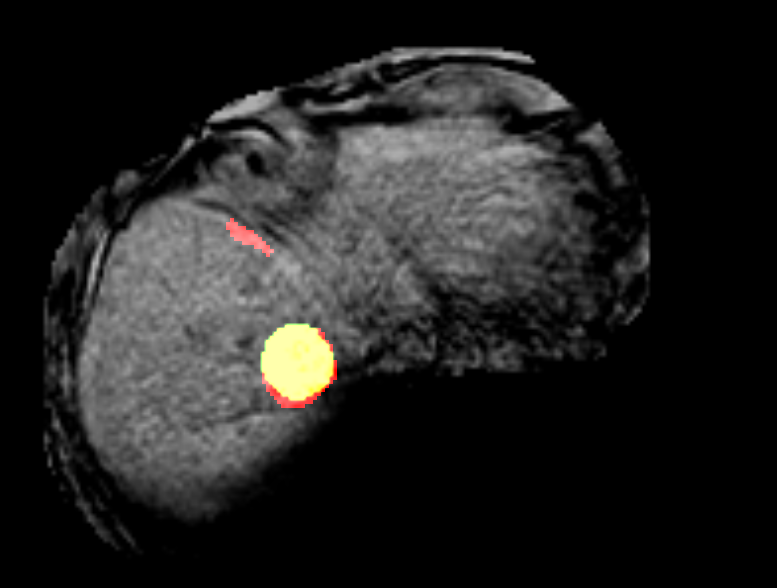}
  \caption{DCE MR image overlayed with true and predicted lesion masks. The red prediction corresponds to a false positve detection, while the ground truth annotation is shown in yellow.}
\end{subfigure}\hspace{0.05\textwidth}%
\begin{subfigure}[t]{0.3\linewidth}
  \includegraphics[height=4.0cm]{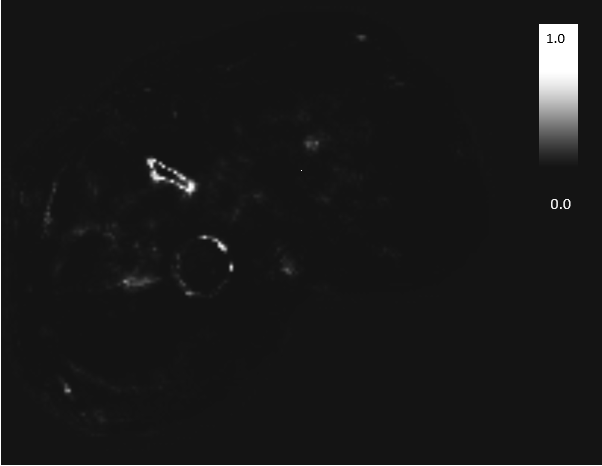}
  \caption{Uncertainty map containing per-voxel entropy computed from the mean softmax prediction.}\medskip
\end{subfigure}
\caption{Lesion detection and uncertainty quantification}
\label{fig:unc_illus}
\end{figure*}

The network is trained using 2-D slices from the 3-D DCE and DWI MR images. As described in ~\cite{gal_dropout_2015}, the network is trained with dropout and at test-time, outputs obtained from multiple passes through the same network (with dropout enabled) are used to estimate model uncertainty. Each pass can be thought of as a sample from the weight posterior distribution and averaging the outputs can be thought of as marginalizing out the weight posterior to obtain an estimate of the model likelihood. Thus, the mean output over multiple passes is taken to be the final prediction. To quantify the uncertainty, we calculate the entropy of the mean softmax prediction given by $-\sum_{c=1}^C \hat{p}(c)\text{log}\hat{p}(c)$, where $c$ is class index and $\hat{p}(c)$ is the softmax value for that class. We create the binary prediction by thresholding the mean softmax output at 0.5. To remove noisy detections and fill small holes, this step is followed by a post-processing step involving binary closing with a $3 \times 3 \times 3$ structuring element and opening with a $3 \times 3$ plus-shaped structuring element. An example of a lesion prediction and associated uncertainty map is shown in Figure \ref{fig:unc_illus}. 

During training we use the Adam~\cite{kingma_adam:_2015} optimizer with an initial learning rate of $10^{-3}$ in combination with the PolyLR scheduler~\cite{chen_deeplab_2017} to decrease the learning rate as training progresses. We extract 128x128 overlapping patches from the 256x256 size image slices (5 per image) and feed this to the neural network. We use rotations using angles sampled from a uniform distribution over $[-45, 45]$ to augment the training images. We use a weighted version of the standard cross-entropy loss to address the class-imbalance. To estimate uncertainty during test-time, we use 20 forward passes for each image patch. We decided on 20 forward passes since we saw no improvement in segmentation performance or false positive classification on the validation set on increasing the number of passes. We train the neural networks for 120K iterations.

\subsection{Feature Extraction and Classification}

\begin{figure}[htb]
  \centering 
  \includegraphics[width=8.5cm]{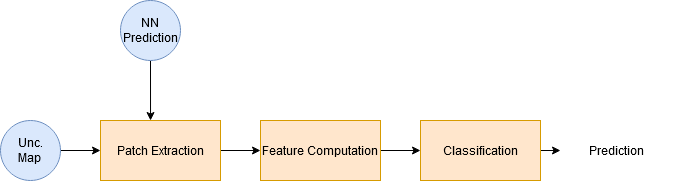}
  \caption{Feature extraction and lesion classification pipeline}
  \label{fig:les_class}
\end{figure}

The feature extraction and classification pipeline is shown in Figure \ref{fig:les_class}. For each patient, we extract 3-D patches from the uncertainty map corresponding to regions in the neural network output where lesions have been detected. In our analysis we found that false positive predictions tended to have a smaller volume as compared to true positive predictions. Therefore, in addition to the mean uncertainty, we selected the maximum diameter of the detection as a feature.

Additionally, we automatically selected features from a set of 107 features extracted using PyRadiomics\footnote{https://github.com/Radiomics/pyradiomics}. Linear models with L1 penalty produce sparse classifiers with many of the feature coefficients set to zero after training. We use such a classifier trained on the validation patient dataset to select the top-2 features with non-zero coefficients for each configuration. We compare the performance of these automatically selected features with features we chose manually.

We use a support vector machine (SVM) to classify the feature vector as a true or a false positive lesion. The classifier is trained using patches extracted from uncertainty maps computed for patients part of the neural network validation set (5 patients). Hyper-parameters are selected using a grid search in combination with cross-validation over this data.

\section{Results}

In this section we show results of lesion detection and false positive classification for three different configurations:
\begin{itemize}
    \item Baseline ($p=0$, No dropout)
    \item Low dropout ($p=0.3$)
    \item High dropout ($p=0.5$)
\end{itemize}

We vary the dropout rate ($p$) to analyze the behavior of MC-Dropout over a range of values. Using a dropout rate higher than 0.5 lead to unstable training. For each configuration, we train 5 different neural network instances, each of which has a different train-validation data split. The set of test patients used to report the performance are the same across all runs and configurations. 

\subsection{Lesion detection}

The results for lesion detection are shown in Table \ref{tab:nn_preds}. A connected region in the neural network prediction is counted as a single lesion prediction. If such a prediction has a non-zero overlap with ground truth annotation, it is considered \emph{detected} i.e. a true positive. If there is no overlap, then that lesion is a false positive.

We see that increasing the dropout rate reduces the number of false positives which could be attributed to its regularizing effect and slight improvement in calibration~\cite{jungo_analyzing_2020}. 

\begin{table}[htbp]
\resizebox{0.4\textwidth}{!}{%
\begin{tabular}{|l|l|l|l|}
\hline
\textbf{Configuration} &
  \textbf{\begin{tabular}[c]{@{}l@{}}True Positive\\ Predictions\end{tabular}} &
  \textbf{\begin{tabular}[c]{@{}l@{}}False Positive\\ Predictions\end{tabular}}\\ \hline
Baseline     & 75 & 41.4 \\ \hline
Low Dropout  & 75 & 38.6 \\ \hline
High Dropout & 75 & 29.0 \\ \hline
\end{tabular}%
}
\caption{Total number of true and false positive predictions by the neural network in the test-set. Mean taken over results of 5 separately trained neural networks per configuration.}
\label{tab:nn_preds}
\end{table}

\subsection{False positive classification}
\label{sec:fpc}

In Table \ref{tab:feat_cv}, we show the cross-validation results for the SVM training data (mean and standard deviation) for all configurations using manual and automatic feature selection. In all cases, the cross-validation accuracy for the manually selected features is higher than or equal to that of the automatically selected features. This led us to choose the manually selected features to perform classification and report results on the test set. 

\begin{table}[htpb]
\resizebox{0.5\textwidth}{!}{%
\begin{tabular}{|l|l|l|}
\hline
\textbf{Configuration} & \textbf{Manual Feature Selection} & \textbf{Automatic Feature Selection} \\ \hline
Baseline     & $0.996 \pm 0.008$ & $0.989 \pm 0.012$ \\ \hline
Low Dropout  & $0.996 \pm 0.008$ & $0.996 \pm 0.008$ \\ \hline
High Dropout & $0.984 \pm 0.032$ & $0.974 \pm 0.044$ \\ \hline
\end{tabular}%
}
\caption{Cross-validation accuracy (mean and standard deviation) for manual and automatic feature selection }
\label{tab:feat_cv}
\end{table}

\begin{table}[htpb]
\resizebox{0.5\textwidth}{!}{%
\begin{tabular}{|l|l|l|l|l|}
\hline
\textbf{Model} & \textbf{Accuracy $(\uparrow)$} & \textbf{Sensitivity $(\uparrow)$} & \textbf{Specificity $(\uparrow)$} & \textbf{F1-Score $(\uparrow)$} \\ \hline
Baseline     & $0.927 \pm 0.044$ & $0.791 \pm 0.228$ & $0.970 \pm 0.053$ & $0.845 \pm 0.169$ \\ \hline
Low dropout  & $0.962 \pm 0.011$ & $0.914 \pm 0.049$ & $0.970 \pm 0.019$ & $0.911 \pm 0.048$ \\ \hline
High dropout & $0.957 \pm 0.025$ & $0.902 \pm 0.069$ & $0.984 \pm 0.010$ & $0.928 \pm 0.040$ \\ \hline
\end{tabular}%
}
\caption{False positive classification metrics for test patients using manually selected features}
\label{tab:fp_results}
\end{table}

In Table \ref{tab:fp_results} we report classification metrics for the false positive detection task. We see that the dropout configurations have a better accuracy and sensitivity i.e. they are much better at classifying false positive predictions made by the neural network correctly.

The specificity metric tells us the ability of the classifier to correctly classify a true positive lesion. On this metric, the high dropout configuration mis-classifies around $1.6 \%$ of true lesions, the best among the 3 configurations. 

In Table \ref{tab:fp_stats} we show the total number of predictions (true and false positives) in the test set before and after the feature based classification. The number of false positives is smallest for the high dropout configuration after classification.  Additionally, it retains the most number of true positives owing to its better specificity. 

\begin{table}[htb]
\resizebox{0.5\textwidth}{!}{%
\begin{tabular}{|l|l|l|l|l|}
\hline
\textbf{Configuration} &
  \textbf{\begin{tabular}[c]{@{}l@{}}True Positives\\ (Before)\end{tabular}} &
  \textbf{\begin{tabular}[c]{@{}l@{}}True Positives\\ (After)\end{tabular}} &
  \textbf{\begin{tabular}[c]{@{}l@{}}False Positives\\ (Before)\end{tabular}} &
  \textbf{\begin{tabular}[c]{@{}l@{}}False Positives\\ (After)\end{tabular}} \\ \hline
Baseline     & 75 & 72.8 & 41.4 & 8.9 \\ \hline
Low Dropout  & 75 & 73   & 38.6 & 3.3 \\ \hline
High Dropout & 75 & 73.8 & 29.0 & 2.7 \\ \hline
\end{tabular}%
}
\caption{Total number of true and false positive predictions in test-set (mean over runs) before and after classification. }
\label{tab:fp_stats}
\end{table}

%
%

\section{Discussion and Conclusion}

Our results show that the neural network with a dropout rate of $0.5$ filters out close to $90$\% of false positive detections in the neural network output. 
By choosing MC-Dropout to estimate uncertainty, we consider only the uncertainty inherent in the model and not the data. The method might be further improved by combining MC-Dropout with techniques to estimate data uncertainty.

We could not use this approach to correct false negatives. These were extremely small in size and would get filtered out during post-processing of the predicted mask as noise.

Using more than two features did not improve the performance of the false positive classification. Further investigation into the robustness of the manually selected features over image modality, organ, uncertainty estimation technique is required.

\section{Compliance with Ethical Standards}
The UMCU Medical Ethical Committee has reviewed this study and informed consent was waived due to its retrospective nature.

\section{Acknowledgements}

This work was financially supported by the project IMPACT (Intelligence based iMprovement of Personalized treatment And Clinical workflow supporT) in the framework of the EU research programme ITEA3 (Information Technology for European Advancement). The authors declare no conflict of interest. 
\bibliographystyle{ieeetr}
\bibliography{root}

\end{document}